\documentclass{IEEEtran}
\usepackage{SIunits}
\usepackage{graphicx}
\ifCLASSINFOpdf
\else
\fi
\begin{document}
\title{Planar-Waveguide External Cavity Laser Stabilization for an Optical Link with \( 10^{-19} \) Frequency Stability}
\author{C. Clivati, A. Mura, D. Calonico, F. Levi, G. A. Costanzo, C. E. Calosso and A. Godone
\thanks{C. Clivati is with Politecnico di Torino, Torino, Italia and I.N.Ri.M. - Istituto Nazionale di Ricerca Metrologica - Torino, Italia. e-mail: c.clivati@inrim.it
  		}
\thanks{A. Mura, D. Calonico, F. Levi, C. E. Calosso and A. Godone are with I.N.Ri.M.-Istituto Nazionale di Ricerca Metrologica - Torino, Italia}%
\thanks{G. A. Costanzo is with Politecnico di Torino, Torino, Italia
  		}%
\thanks{This work has been funded by Compagnia di S. Paolo (2007.2821)}}

\maketitle

\begin{abstract}
We stabilized the frequency of a compact planar-waveguide external cavity laser (ECL) on a Fabry-P\'erot cavity (FPC) through a Pound-Drever-Hall scheme. The residual frequency stability of the ECL is \(10^{-14} \), comparable to the stability achievable with a fiber laser (FL) locked to a FPC through the same scheme. We set up an optical link of \unit{100}{\kilo\metre}, based on fiber spools, that reaches \( 10^{-19} \) relative stability, and we show that its performances using the ECL or FL are comparable. Thus ECLs could serve as an excellent replacement for FLs in optical links where cost-effectiveness and robustness are important considerations. 
\end{abstract}
\IEEEpeerreviewmaketitle
\section{Introduction}
\IEEEPARstart{U}{ltra} stable lasers with narrow linewidth are an enabling technology for a variety of scientific applications. They allow to excite ultra narrow atomic transitions in high-resolution spectroscopy and in optical atomic clocks \cite{roseband, ludlow2}.
They are also required in physics experiments such as tests of fundamental constants variations (for a review, see \cite{karsenboim}) or search of deviations from the Standard Model \cite{herman}. In addition, highly coherent lasers are used to realize frequency transfer systems based on optical fibers \cite{ma, jiang}. This is a fundamental topic in primary metrology, since it allows frequency dissemination and remote clocks comparisons with the  highest resolution available at present. Traditional satellite techniques like GPS carrier-phase \cite{gpscp}, Two-Way Satellite Time and Frequency Transfer (TWSTFT) via geostationary satellites \cite{twstft} and GPS Precise Point Positioning (PPP) \cite{ppp}, have a relative stability of \( 10^{-14}/\text{day} \)~\cite{bauch}. Therefore, they are no longer able to withstand the performances of the new generation of optical frequency standards. On the other hand, the frequency transfer on compensated optical fibers can reach a stability of \(  10^{-15}\tau^{-1}  \) over hauls of several hundreds kilometres \cite{predehl, williams, lopez}. These results point out a nearly \(10^5 \)  fold improvement in relative stability at \unit{1000}{\second} with respect to satellite techniques.

The development of optical frequency links led to a number of new technological issues related to the realization of ultra stable laser sources, that require balancing high performances with compactness, robustness and low cost.\\

The stabilization of the laser source to the sub-hertz level is typically accomplished by locking a fiber laser (FL) to a high finesse Fabry-Perot cavity (FPC) using the Pound-Drever-Hall technique \cite{drever}. Examples of locking to long fiber spools are reported as well \cite{kefelian} even if this technique is at a preliminary stage. However, not only the cavity must be stabilized and insulated from thermal and mechanical noise, but the  linewidth of the free-running laser needs to be substantially narrower than achievable control bandwidths of \( \sim \unit{1}{\mega\hertz}\). The stabilization on a FPC allows linewidth reduction from a few kilohertz to the sub-hertz level. The residual stability  is eventually limited by the thermal fluctuations of the cavity \cite{numata} and achieves some parts in \( 10^{-15} \) in terms of relative Allan deviation \cite{leibrandt}.\\
Distributed-Bragg-Reflector (DBR) FLs are preferred, especially in fiber coupled systems, thanks to their broad wavelength tunability and to the relatively low frequency noise level. However, planar-waveguide external cavity lasers (ECLs) are becoming a promising option as well. Their performances are similar to the ones of FLs; in addition, they are cheaper and more agile than FLs. These qualities make them attractive for the realization of robust optical links and transportable optical clocks.\\

In this paper we demonstrate the stabilization of a compact ECL on a FPC through a Pound-Drever-Hall scheme and compare its performances to that of a FL.\\
 We built two identical stabilization setups to compare the beat note between two FLs and between a FL and the ECL. \\
Then, we used the FPC-stabilized ECL to set up an optical link over a \unit{100}{\kilo\metre} fiber spool inside our laboratory, to show a possible application. We demonstrated performances of the laser beyond the level first reported by  Numata \emph{et al.} in \cite{numataOrion}, who stabilized an ECL on acetylene (\(^{13}\text{C}_2\text{H}_2 \)) at \(10^{-13} \) level of Allan deviation. This paper shows that the ECL under test is an effective alternative to FLs for the realization of compact ultra stable laser sources for high accuracy metrology and for the realization of coherent optical links.

\section{Ultra-stable laser source realization}
\subsection{Experimental setup}
\begin{figure}[!t]
\centering
\includegraphics[width=3.5in]{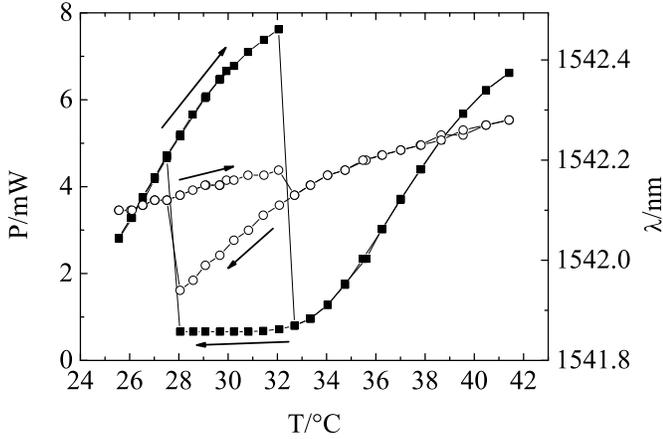}
\caption{The hysteresis curves of power (filled squares, left \(y \) axis)  and wavelength (empty circles, right \(y \) axis) vs temperature.}
\label{fig:hysteresis}
\end{figure}

\begin{figure}[!t]
\centering
\includegraphics[width=3.5in]{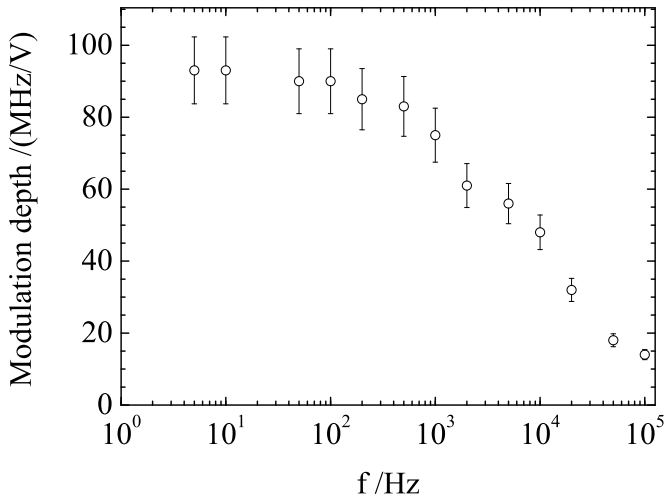}
\caption{Modulation depth of the ORION module with \unit{1}{\volt} modulation voltage vs the modulation frequency.}
\label{fig:mod_depth}
\end{figure}

\begin{figure}[!t]
\centering
\includegraphics[width=3.5in]{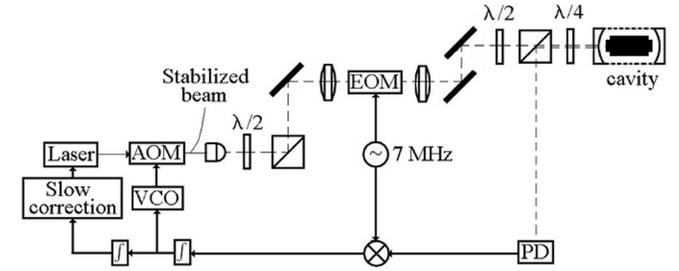}
\caption{The Pound-Drever-Hall stabilization system. Thin solid lines indicate fibered paths, thin dashed lines indicate free-air paths; thick lines indicate electronic paths. VCO is the Voltage Controlled Oscillator; \( \lambda/2\) and \( \lambda/4\) are half and quarter-wave plates.}
\label{fig:PDH_scheme}
\end{figure}

We tested a laser module called ORION, built by Redfern Integrated Optics (RIO). It includes a butterfly-packaged planar-waveguide fiber-coupled ECL (PLANEX\textsuperscript{TM}). The size of the module is about \unit{10}{\centi\metre} \(\times \) \unit{6}{\centi\metre} \( \times \) \unit{1}{\centi\metre}, that is about \unit{0.06}{\litre}. On the other hand, the typical size of a FL is around \unit{10}{\centi\metre} \(\times \) \unit{45}{\centi\metre} \( \times \) \unit{38}{\centi\metre}, that is \unit{17}{\litre}. ORION is operating at \unit{1542}{\nano\metre} and can be frequency tuned either through the injection current or through the temperature. The temperature setpoint is changed via a RS232 serial communication port. As reported in Fig. \ref{fig:hysteresis}, the laser exhibits hysteresis in power and wavelength vs laser temperature. However, during stable operation, the temperature tuning is not used and we never observe mode hops or abrupt changes in the emitted power.\\
The module can provide a power up to \unit{8}{\milli\watt}.
The Current modulation is easily obtained applying a voltage to the module, in the range \( \pm \unit{2.8}{\volt} \). The modulation performance is evaluated by the beat note between ORION and a FL locked to a FPC. 
For three different amplitudes (\unit{20}{\milli \volt} ,\unit{180}{\milli \volt}, \unit{1}{\volt}) we measured the same modulation amplitude of \(\sim \unit{90}{\mega\hertz\per\volt}\) and a \unit{-3}{\deci\bel} bandwidth of \unit{3}{\kilo\hertz}. 
Fig. \ref{fig:mod_depth} shows the modulation amplitude with \unit{1}{\volt} modulation voltage.\\
This rather low bandwidth is not a practical limitation in the stabilization scheme, because an Acousto-Optic Modulator (AOM) is used to provide a high-bandwidth feedback, whereas the drift of the laser is sufficiently small that the range of the current port is adequate to keep it locked.\\
Once the proper operating temperature is set, the ORION module stays locked to the FPC for extended periods of time without further adjustment.\\
Fig. \ref{fig:PDH_scheme} shows the frequency stabilization scheme: we realized two independent systems, to characterize the performances by means of the beat note between indipendently stabilized lasers.\\
A \unit{40}{\mega\hertz} AOM is used to compensate the laser noise and we use a resonant Electro-Optic Modulator (EOM) at \unit{7}{\mega\hertz} for Pound-Drever-Hall locking. 
The control loop is a Proportional-Integral-Derivative circuit with nearly \unit{100}{\kilo\hertz} bandwidth; the  bandwidth of the plant is limited by the delay time of our AOM, that is \(\sim \unit{1}{\micro\second}\). Low frequency noise up to a bandwidth of \unit{1}{\kilo\hertz} and drift are corrected by the current modulation in the ORION module and by a piezoelectric actuator in the FL.\\
The reference for the laser stabilization of each circuit is an independent Corning ULE\textsuperscript\textregistered cavity with finesse of \(120\,000\), and Free Spectral Range of \unit{1.5}{\giga \hertz}. The cavities are notched in order to reduce the sensitivity of the optical axis length to seismic noise \cite{chen};  
we measured a sensitivity of the optical axis length to gravitational force of \(\unit{4.5}{\pico\metre}/\unit{(}{\metre\per\second\squared}) \), corresponding to \(\unit{9}{\kilo\hertz}/\unit{(}{\metre\per\second\squared}) \).\\
Each cavity is housed in a copper thermal shield inside a vacuum chamber at \(6 \times 10^{-6} \unit{}{\pascal} \) and
the whole vacuum system is wrapped in multiple layers of foam rubber to insulate from environmental thermal fluctuations. A digital control loop compensates the temperature variations on the copper shield with a bandwidth of some millihertz, driving two heaters on the outside of the vacuum vessel. We achieve a thermal stability of \unit{1}{\micro\kelvin} at \unit{1}{\second}, whereas on the long term we observe a peak to peak variation of \unit{1}{\milli\kelvin}.\\
The laser power impinging the FPC is less than \unit{20}{\micro\watt}, and kept as low as possible to reduce etalon effects and amplitude-to-frequency noise conversions inside the cavity. We measured relative power fluctuations of \(5 \times 10^{-4}\) at \unit{1}{\second} and a power to frequency conversion coefficient of about \unit{20}{\hertz\per\micro\watt}, resulting in a negligible contribution to the laser stability.   
No significant differences were measured in the amplitude noise power spectrum between the FL and the ORION.

\subsection{Experimental results}

\begin{figure}[!t]
\centering
\includegraphics[width=3.5in]{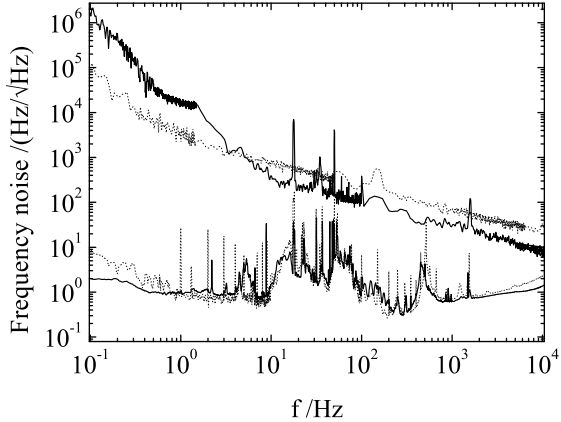}
\caption{Frequency noise of free running ORION (upper dotted line) and FL (upper solid line) and closed loop frequency noise of the beat note between ORION and FL (lower dotted line) and two FLs (lower solid line).}
\label{fig:laser_noise}
\end{figure}

\begin{figure}[!t]
\centering
\includegraphics[width=3.5in]{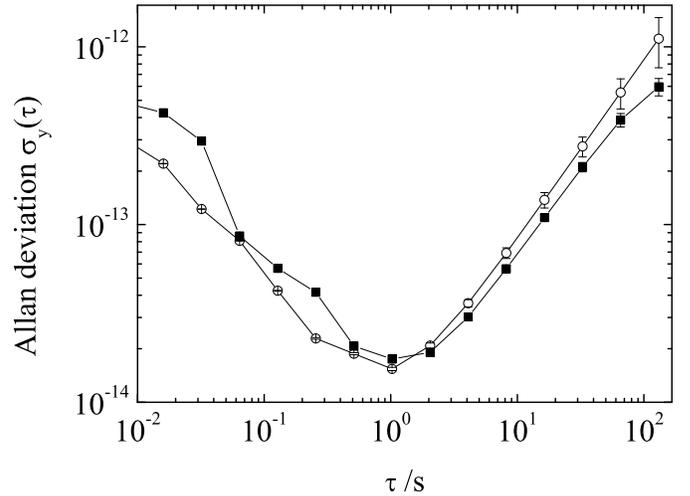}
\caption{Allan deviation of the beat note between ORION and FL (filled squares) and two FLs (empty circles).}
\label{fig:allan_dev_beatnote}
\end{figure}

To characterize the noise of the free running lasers we locked them to the FPC, using only the fast feedback provided by the AOM; we then acquired the correction signal sent to the Voltage Controlled Oscillator (VCO) with a Fast Fourier Transform (FFT) Spectrum Analyser. This voltage is proportional to the free running frequency noise of the laser, up to the control bandwidth, since it is expected to be greater than the noise of the reference cavity. 
The results are plotted in Fig. \ref{fig:laser_noise}.\\ 
A comparison of the frequency noise spectrum of the FL and the ORION shows a higher frequency noise slope of the FL below \unit{10}{\hertz}, maybe caused by mechanical and acoustic noise in the fiber, while the FL noise is lower above \unit{30}{\hertz}. Also, the FL shows to be more sensitive to acoustic resonances between \unit{20}{\hertz} and \unit{100}{\hertz}.\\
Following the approach of Di Domenico \emph{et al}. \cite{didomenico} the open loop frequency noise spectrum has been integrated from \unit{10}{\hertz} to \unit{2}{\kilo\hertz} (\unit{3}{\kilo\hertz} for the ORION noise) to evaluate the free running laser fast linewidth. We choosed \unit{10}{\hertz} as lower integration limit to prevent the divergence of the laser linewidth at lower frequencies. The upper integration limit was chosen because higher frequencies only contribute to the wings of the line shape and not to the laser linewidth. We used this approach since it allows to exclude the contribution of environmental acoustic peaks from the evaluation of the laser linewidth. We deduced a laser linewidth of \unit{5}{\kilo\hertz} for our free-running FL and a linewidth of \unit{17}{\kilo\hertz} for the ORION module. \\
To characterize the noise of the locked laser, we downconverted the beat note (\( \sim \unit{300}{\mega\hertz}\)) between two FLs and a FL and the ORION, and processed it with a digital phasemeter. Since the stabilization scheme is the same and the free running noise of the two lasers is similar, we expected a similar performance of the stabilized ORION and the FL. 
Fig. \ref{fig:laser_noise} reports the residual frequency noise spectra of the beat note between two FLs and between ORION and a FL. As expected, the free running noise of the ORION module is not a practical limitation to the frequency stabilization, since its contribution can be kept negligible with respect to the residual noise of the cavities.\\
The beat note is strongly modulated by the peaks clearly visible on the frequency noise spectrum between \unit{10}{\hertz} and \unit{100}{\hertz}: they arise from acoustic and seismic vibrations of the cavities and of the lasers and may be reduced in the future through a better insulation.
\\
In Fig. \ref{fig:allan_dev_beatnote} we show the relative Allan deviation of the beat note between two FLs and between ORION and a FL.
In both situations we reach an Allan deviation of \(2 \times 10^{-14}\) at \unit{1}{\second}; as mentioned before, we attribute such residual instability to the cavity itself and not to the laser source. The relatively large drift (about \unit{2.5}{\hertz\per\second}) is consistent with the residual temperature drift of the cavities.

\section{The optical link}
\subsection{Experimental setup}
\begin{figure}[!t]
\centering
\includegraphics[width=3.5in]{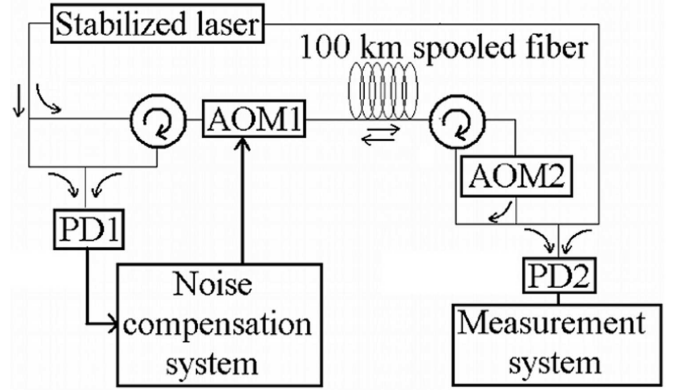}
\caption{Scheme of the optical link. PD1 and PD2 indicate the photodiodes. Thin lines indicate fibered paths, thick lines indicate electronic paths.}
\label{fig:scheme_link}
\end{figure}

We used the stabilized ORION to set up an optical link over a \unit{100}{\kilo\metre} fiber spool inside our laboratory. The scheme is shown in Fig. \ref{fig:scheme_link}.
The stabilized laser is sent through \unit{100}{\kilo\metre} of fiber spool to the remote end of the link and there it is reflected back to the local end to actively compensate the phase noise generated by the link. To this purpose two AOMs are placed on the fiber path. The first one (AOM1) is used to actively compensate the noise of the link, the second (AOM2) is used at the remote end to frequency shift the retro-reflected light, in order to distinguish spurious reflections from the round-trip signal. The retro-reflected light is compared to the input on photodiode PD1, to extract the phase noise of the link and to generate the correction signal. At the remote end, the transmitted radiation is compared to the input laser on photodiode PD2 to measure the link performance.\\ 
The system is enclosed in an insulation box to reduce temperature fluctuations and acoustic noise on the fibers.\\
\subsection{Experimental results}
\begin{figure}[!t]
\centering
\includegraphics[width=3.5in]{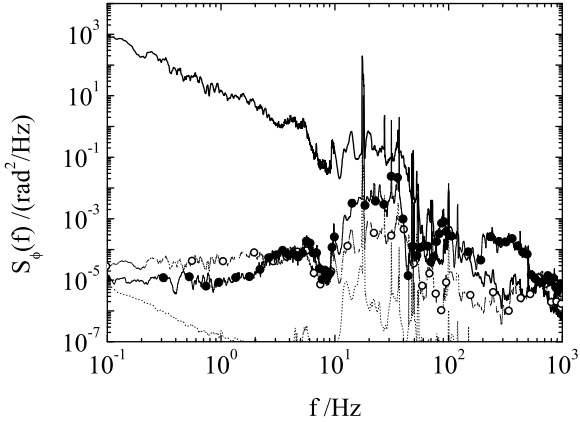}
\caption{Phase noise of the optical link at the remote end, using the ORION as a light source. Solid line: free running link. Solid line, filled circles: compensated link. Dotted line, empty circles: expected delay-limited residual noise. Dotted line: system noise floor.}
\label{fig:link_spiego}
\end{figure}

\begin{figure}[!t]
\centering
\includegraphics[width=3.5in]{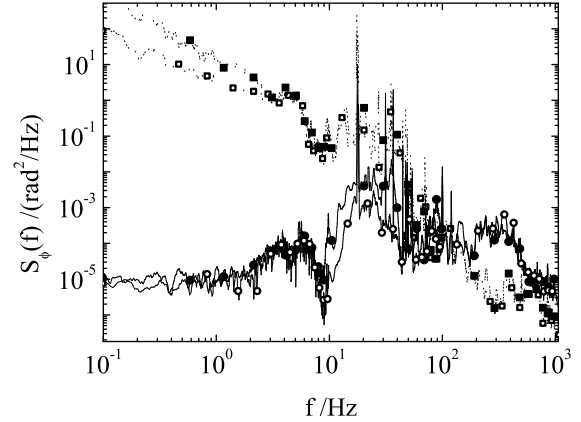}
\caption{Comparison of free running and compensated link performances using ORION or the FL. Dotted line, filled squares: free running link with ORION; dotted line, empty squares: free running link with FL. Solid line, filled circles: compensated link with ORION; solid line, empty circles: compensated link with FL.}
\label{fig:link_comparison}
\end{figure}

\begin{figure}[!t]
\centering
\includegraphics[width=3.5in]{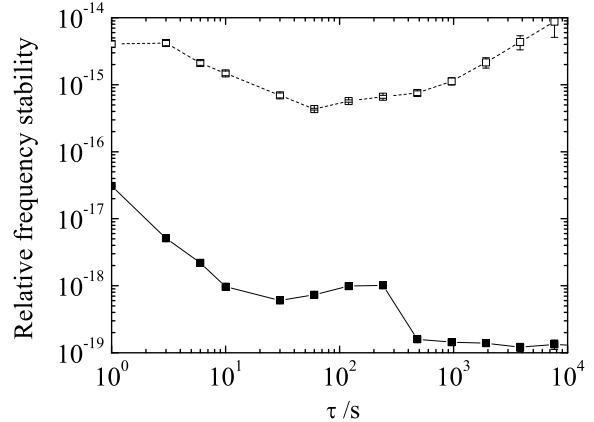}
\caption{Relative stability of the free running (dashed line, empty squares) and compensated link using the ORION (solid line, filled squares) as a light source, aquired with a \(\Lambda\)-type frequency counter. 
}
\label{fig:allan_dev_link}
\end{figure}

We measured stability and the phase noise spectrum of the beat note between injected and transmitted radiation; 
phase noise spectra are acquired through a digital phasemeter, whereas link stability over longer periods is measured by counting the downconverted beat note with a high-resolution frequency counter. Since this is a \(\Lambda\)-type frequency counter, it leads to a quantity similar to the modified Allan deviation; a detailed discussion and scaling factor can be found in \cite{rubiola, lambdaType}.\\
Fig. \ref{fig:link_spiego} reports the phase noise spectra at open and closed loop (using the ORION as laser source). The fiber noise cannot be completely removed at the remote end even for infinite loop gain, because we are locking the link at the local end. The closed loop phase noise at the remote end (solid line, filled circles) approaches to the expected delay-limited noise (dotted line, empty circles) at frequencies between \unit{0.1}{\hertz} and \unit{20}{\hertz}: as evaluated in \cite{williams}, assuming that the noise is uncorrelated over the link, the minimum residual phase noise is \( \frac{1}{3}(2 \pi f \tau_\text{F})^2 \Phi^2_\text{F}(f) \)
where  \( \Phi^2_\text{F}(f) \) represents the phase noise of the fiber and \( \tau_\text{F}\) is the fiber delay.\\
At frequencies higher than \unit{20}{\hertz}, the phase noise of the compensated link is limited by the Phase Locking Loop (PLL) bandwidth; at frequencies lower than \unit{0.1}{\hertz} the compensation is limited by the noise floor (dotted line) of our system.\\
In Fig. \ref{fig:link_comparison}, we compare the phase noise of the free running and compensated link using ORION (respectively dotted line with filled squares and solid line with filled circles) and FL (respectively dotted line with empty squares and solid line with empty circles); as mentioned before, the link noise is limited by the delay and by the PLL gain inside the loop bandwidth. 
As expected, the compensated link shows quite similar performances with ORION and the FL; the small differences  at frequencies around \unit{30}{\hertz} can be due to non stationary acoustic and seismic noise in the laboratory.\\
Fig. \ref{fig:allan_dev_link} reports the relative stability of the optical link using ORION as a laser source; 
it exhibits a \( \tau^{-3/2}\) dependence at short integration times, as expected when acquiring white phase noise with a \( \Lambda\)-type frequency counter. The bump around \unit{250}{\second} corresponds to half cycle of the air conditioning, whereas the ultimate stability reaches the \( 10^{-19}\) level, limited by the temperature fluctuations on the uncompensated fibers. The unmatched fiber paths would need to be further reduced in order to limit the thermal expansion and polarization changes. \\

\section{Conclusion}
We demonstrated that the ORION module by RIO is suitable to realize an ultra stable laser source by locking it to a high-finesse FPC through the Pound-Drever-Hall technique. High frequency noise is suppressed through a fiber AOM; the laser current is used to increase the loop gain at low frequencies. This laser 
achieves a relative stability of \( 10^{-14}\) at \unit{1}{\second}. The stabilized laser is used to realize an optical link over a length \( L = \unit{100}{\kilo\metre} \) on a fiber spool with an ultimate relative stability of \( 10^{-19} \). The experimental results point out that the ORION module is appropriate for the intended use.
Since the expected delay-limited residual phase noise scales as \(L^3\) while the influence of the laser noise scales as \(L^2\) \cite{williams}, we expect that even for longer links the latter will not constitute a limitation. \\
The compactness and robustness of the ORION module make it an effective alternative to FLs for the realization of ultra stable laser sources and for optical links.

\ifCLASSOPTIONcaptionsoff
  \newpage
\fi

\begin{IEEEbiographynophoto}{Cecilia Clivati} was born in Gattinara, Italy, in 1986. She received the degree in physics in 2010. Presently is a Ph.D. student at Politecnico di Torino and works on the realization of a low phase noise optical frequency link and ultra stable laser sources at the Optic division of Istituto Nazionale di Ricerca Metrologica - INRIM, Torino, Italy.\\
\end{IEEEbiographynophoto}

\begin{IEEEbiographynophoto}{Alberto Mura} was born in Sassari, Italy, in 1972. He graduated in electronic engineering in 2005. In 2009, he received his Ph.D. degree in metrology with a dissertation on the measurement of the gravitational constant. At the moment is with INRIM and works on the realization of a low phase noise optical frequency link and ultra stable laser sources.\\
\end{IEEEbiographynophoto}

\begin{IEEEbiographynophoto}{Davide Calonico} was born in Torino in 1975. He received his Ph.D. degree in metrology at Politecnico di Torino in 2003. During his Ph.D. program he worked on a laser-cooled \(^{133}\)Cs fountain frequency
standard at IEN (now INRIM) in Turin as well as on a dual \(^{133}\)Cs-\(^{87}\)Rb fountain at BNM-SYRTE in Paris in collaboration with Ecole Normale Sup\'erieure. Now he is a researcher at INRIM, where he is involved with laser-cooled atomic frequency standards in the microwave and visible frequency domain.\\
\end{IEEEbiographynophoto}

\begin{IEEEbiographynophoto}{Filippo Levi} received the degree in physics from the University of Torino in 1992. In 1996, he received 
the Ph.D. degree in metrology from the Politecnico di Torino. Since 1995, he has been a researcher at 
the Time and Frequency division of IEN (now INRIM), Torino, Italy, where he is responsible of the realization of the Italian Cesium fountain primary frequency standard. His other 
main research field is the study of cell frequency standards and, in particular, the study of the coherent population trapping (CPT) phenomena. In 1998 and then in 2000 and 2001, he was guest researcher at NIST for studies on the application of cooling techniques to atomic frequency  standards. His research activity is currently concerned with the realization of atomic frequency standards in the microwave region and the development of laser-cooled frequency standards. He received the European Frequency and Time young scientist award in 1999.\\
\end{IEEEbiographynophoto}

\begin{IEEEbiographynophoto}{Claudio E. Calosso} was born in  Asti in 1973. He received the degree in engineering from Politecnico di Torino, Italy, in 1998. In 2001, he was a guest researcher at NIST for studies on multi-launch atomic fountains. In 2002, he received the Ph.D degree in communication and electronic engineering from the Politecnico di Torino. His research activity was devoted to the development of the electronics for the atomic fountain and for the CPT maser. He is now with the Optic division of the INRIM of Turin. His research activity concerns the development of an optically pumped frequency standards.\\
\end{IEEEbiographynophoto}

\begin{IEEEbiographynophoto}{Giovanni A. Costanzo} was born  in S. Benedetto del Tronto, Italy in 1964. He graduated  in electronic engineering from  the University of  Ancona, Italy in 1989. In  1995, he  received the Ph.D. in metrology  from Politecnico di Torino with  a dissertation on  the development and evaluation of the high C-field Cs beam frequency standard. He then went to NRLM (Tsukuba, Japan) as a postdoctoral fellow, to work on the realization of  the laser  system necessary for  the Cs fountain. In  1998 he  joined, as a permanent  staff member, the Dipartimento di Elettronica  at  the Politecnico di  Torino. 
Presently, he works on atomic frequency standards. \\
\end{IEEEbiographynophoto}

\begin{IEEEbiographynophoto}{Aldo Godone} received the Dr. Ing. degree in electronic engineering 
from the Politecnico di Torino, Italy. In 1974 he joined the Time and Frequency department of the IEN Galileo Ferraris (now INRIM), Torino, Italy, where he is involved in the development of atomic frequency 
standards in the sub-millimeter and microwave regions. In collaboration with PTB, between 1977 and 1987, he studied and developed new techniques for the realization of low-noise frequency multiplication chains, with particular interest in the propagation of the phase noise in the multiplication process. Between 1980 and 1990, he developed the Mg beam frequency standard in the \unit{600}{\giga\hertz} region. His research activity 
also includes the development of techniques for the extension of high-resolution frequency measurements up to the infrared region, and since 1990, the development of highly stable frequency standards. Currently his main research interest is in the realization of laser pumped based cell frequency standards and in the realization and maintenance of primary frequency standards. For his fundamental contributions to the development of chains of frequency synthesis and frequency standards in the far-infrared and to the field of atomic clocks based on optical pumping, Dr. Godone received in 2008 the European Frequency and Time award.
\end{IEEEbiographynophoto}


\begin{thebibliography}{1}
\bibitem{roseband}{T. Roseband et al., ``Frequency ratio of Al\(^+\) and Hg\(^+\) single-ion optical clocks; metrology at the 17\(^\text{th}\) decimal place,'' \emph{Science}, vol. 319, pp. 1808-1812, Mar. 2008.}\\
\bibitem{ludlow2}{A. D. Ludlow et al., ``Sr lattice clock at \(10^{-16}\) fractional uncertainty by remote optical evaluation with a Ca clock", \emph{Science}, vol. 319, pp. 1805-1808, Mar. 2008.}\\
\bibitem{karsenboim}{S. G. Karshenboim, ``Fundamental physical constants: looking from different
angles", \emph{Can. J. Phys.}, vol. 83, pp. 767-811, Aug. 2005.}\\
\bibitem{herman}{S. Hermann et al., ``Rotating optical cavity experiment testing Lorentz
invariance at the \(10^{-17} \) level", \emph{Phys. Rev. D}, vol. 80, pp. 105011 - 105011-8, Nov. 2009.}\\
\bibitem{ma}{L. S. Ma et al., ``Delivering the same optical frequency
at two places: accurate cancellation of phase noise introduced by an optical fiber or other
time-varying path", \emph{Opt. Lett.}, vol. 19, pp. 1777-1779, Nov. 1994.}\\
\bibitem{jiang}{H. Jiang et al., ``Long-distance frequency transfer over an urban fiber link using optical phase stabilization", \emph{J. Opt. Soc. Am. B}, vol. 25, pp. 2029-2035, Nov. 2008.}\\
\bibitem{gpscp}{J. Ray and K. Senior, ``Geodetic techniques for time and frequency comparisons using GPS phase and code measurements,'' \emph{Metrologia}, vol. 42, pp. 215-232, 2005.}\\
\bibitem{twstft}{D. Kirchner, ``Two-way satellite time and frequency transfer (TWSTFT): principle, implementation, and current performance,'' in \emph{Review of Radio Sciences 1996-1999}, Oxford: Oxford University Press, 1999, pp 27-44. }\\
\bibitem{ppp}{J. Kouba and P. Héroux, 2001, ``Precise Point Positioning using IGS orbits and clock products,'' GPS Solutions, vol. 5, pp. 12-28, 2001.}\\
\bibitem{bauch}{A. Bauch et al., ``Comparison between frequency standards in Europe and the USA at the \(10^{-15}\) uncertainty level", \emph{Metrologia}, vol. 43, pp. 109-120, Febr. 2006.}\\
\bibitem{predehl}{H. Schnatz et al., ``A  \unit{900}{\kilo\metre} long optical fiber link for remote comparison of frequency standards", presented in the 5th Joint  conf. Int. Frequency  Control Symp. and the European Frequency and Time Forum, San Francisco, Ca, May 1–5, 2011.}\\
\bibitem{williams}{P. A. Williams et al., ``High stability transfer of an optical frequency over long fiber-optics links", \emph{J. Opt. Soc. Am. B}, vol. 25, pp. 1284-1293, Jul. 2008.}\\
\bibitem{lopez} {O. Lopez et al., ``Cascaded multiplexed optical link on a telecommunication network for frequency dissemination", \emph{Opt. Exp.}, vol. 18, pp. 16849-16857, Aug. 2010. }\\
\bibitem{drever}{R. W. P. Drever et al., ``Laser phase and frequency stabilization using an optical
resonator", \emph{Appl. Phys. B}, vol. 31, pp. 97-105, Jun. 1983. }\\
\bibitem{kefelian}{F. K\'ef\'elian et al., ``Ultralow-frequency-noise stabilization of a laser by
locking to an optical fiber-delay line", \emph{Opt. Lett.}, vol. 34, pp. 914-916, Apr. 2009.}\\
\bibitem{numata}{K. Numata et al.,``Thermal-Noise Limit in the frequency stabilization of lasers with rigid cavities", \emph{Phys. Rev. Lett.}, vol. 93, pp. 250602-250602-4, Dec. 2004.}\\
\bibitem{leibrandt}{D. R. Leibrandt et al., ``Spherical reference cavities for
frequency stabilization of lasers in non-laboratory environments",  \emph{Opt. Exp.}, vol. 19, pp. 3471-3482, Febr. 2011.}\\
\bibitem{numataOrion} {K. Numata et al., ``Performance of planar-waveguide external cavity laser for precision measurements", \emph{Opt. Expr.}, vol. 18, pp. 22781-22788, Oct. 2010.}\\
\bibitem{chen}{L. Chen et al., ``Vibration-induced elastic deformation of Fabry-P\'erot cavities", \emph{Phys. Rev. A}, vol. 74, pp. 053801-053801-13, Nov. 2006}\\
\bibitem{didomenico}{G. Di Domenico et al., ``Simple approach to the relation between laser frequency
noise and laser line shape", \emph{Appl. Opt.}, vol., pp. 4801-4807, Sept. 2010.}\\
\bibitem{rubiola}{E. Rubiola, ``On the measurement of frequency and of its sample variance with high-resolution counters", \emph{Rev. Sci. Instrum.}, vol. 76, pp. 054703 - 054703-6  , May 2005.}\\
\bibitem{lambdaType}{S. T. Dawkins et al, ``Considerations on the measurementof the stability of oscillators with frequency counters", \emph{IEEE Trans. Ultrason. Ferroelectr. Freq. Control}, vol. 54, pp. 918-925, May 2007.}\\


\end{thebibliography}
\end{document}